\documentclass[12pt]{article} 

\renewcommand{\thefootnote}{\fnsymbol{footnote}}
\usepackage{amsbsy,amssymb,latexsym,amsfonts,amsmath}
\usepackage{mathrsfs}
\usepackage{graphicx}
\usepackage{bm}
\usepackage{comment}
\numberwithin{equation}{section}
\newcommand{\bel}[1]{\begin{equation}\label{#1}}                     
\newcommand{\bal}[1]{\begin{eqnarray}\label{#1}}                     
\newcommand{\be}{\begin{equation}}
\newcommand{\ee}{\end{equation}}
\newcommand{\im}{\mathrm{i}}
\newcommand{\ex}{\mathrm{e}}
\newcommand{\de}{\mathrm{d}}

\newcommand{\mat}[1]{\begin{pmatrix} #1 \end{pmatrix}}

\begin{document}
%%%%%%%%%%%%%%%%%%%%%%%%%%%%%%%%%%%%%%%%%%%%%%%%%%%%%
%%%%%%%%%%%%%%%%%%%%%%%%%%%%%%%%%%%%%%%%%%%%%%%%%%%%%
%
% title page
%
%%%%%%%%%%%%%%%%%%%%%%%%%%%%%%%%%%%%%%%%%%%%%%%%%%%%%
\begin{titlepage}
%%%%%%%%%%%%%%%%%%%%%%%%%%%%%%%%%%%%%%%%%%%%%%%%%%%%%
\begin{flushright}
\normalsize
%\filename
~~~~
NITEP 282\\
August 2006 \\
\end{flushright}
%%%%%%%%%%%%%%%%%%%%%%%%%%%%%%%%%%%%%%%%%%%%%%%%%%%%%

\vspace{15pt}

%%%%%%%%%%%%%%%%%%%%%%%%%%%%%%%%%%%%%%%%%%%%%%%%%%%%%
\begin{center}
{\LARGE   Notes on phase structure and non-vanishing $\beta$ functions } \\
{\LARGE  of one-unitary matrix model \\
\vspace{10pt}
\LARGE }
\end{center}
%%%%%%%%%%%%%%%%%%%%%%%%%%%%%%%%%%%%%%%%%%%%%%%%%%%%%

\vspace{23pt}

%%%%%%%%%%%%%%%%%%%%%%%%%%%%%%%%%%%%%%%%%%%%%%%%%%%%%
\begin{center}
{H. Itoyama$^{a,b}$\footnote{e-mail: itoyama@omu.ac.jp},
  and  R. Yoshioka$^{a,b}$\footnote{e-mail: ryoshioka@omu.ac.jp}  }\\
%%%%%%%%%%%%%%%%%%%%%%%%%%%%%%%%%%%%%%%%%%%%%%%%%%%%%

\vspace{18pt}

%%%%%%%%%%%%%%%%%%%%%%%%%%%%%%%%%%%%%%%%%%%%%%%%%%%%%

$^a$ \textit{Nambu Yoichiro Institute of Theoretical and Experimental Physics (NITEP), Osaka Metropolitan University} \\ 

$^b$ \it Osaka Central Advanced Mathematical Institute (OCAMI), Osaka Metropolitan University\\

3-3-138, Sugimoto, Sumiyoshi-ku, Osaka, 558-8585, Japan \\

\end{center}
%%%%%%%%%%%%%%%%%%%%%%%%%%%%%%%%%%%%%%%%%%%%%%%%%%%%%

\vspace{20pt}

\begin{center}
Abstract\\
\end{center}
%%%%%%%%%%%%%%%%%%%%%%%%%%%%%%%%%%%%%%%%%%%%%%%%%%%%%
Unitary matrix models, among other things, play an important role in representing 
the irregular conformal block
 and hence LEEA of some asymptotically free susy gauge theories.
 Here, we develop a method of how to determine the qualitative structure of phase diagram
  by the deformation of  classical potential, 
  and illustrate this by the examples that contain term up to $\cos n\alpha$, $n \leq 3$.
 We point out that a set of $\beta$ functions on critical ``lines" is nowhere a vanishing vector.  
 This generalizes the $n =1$ GWW case and tells us the third order 
  (rather than second order) phase transition
 in conformity with the range of values for the susceptibility exponent.

%%%%%%%%%%%%%%%%%%%%%%%%%%%%%%%%%%%%%%%%%%%%%%%%%%%%%

\vfill

\end{titlepage}

%%%%%%%%%%%%%%%%%%%%
\renewcommand{\thefootnote}{\arabic{footnote}}
\setcounter{footnote}{0}
%%%%%%%%%%%%%%%%%%%%

%%%%%%%%%%%%%%%%%%%%%%%%%%%%%%%%%%%%%%%%%%%%%%%%%%%%
%%%%%%%%%%%%%%%%%%%%%%%%%%%%%%%%%%%%%%%%%%%%%%%%%%%%
\section{introduction}
      Uses of unitary matrix models are old \cite{Meht2004} and by now extend into several directions. 
    In the series of our continuing investigations 
    \cite{IO1003,MMS1001,IOYone1008,IOYano1805,IOYano1812,IOYano2019,IOYano1909,IOYano2103,
    IOYosh2210,IOYosh2212,CIY2402,IYosh2411}, 
    the relevance of unitary matrix models has arisen from 
     the limiting procedure where some of the flavour masses are sent to infinity 
     from the original hermitian ones (2d Virasoro/W blocks)
      to obtain asymptotically free theories. 
This, of course, lies in the longstanding development \cite{SW9407,Nekr0206,AGT0906}
   of the exact determination of LEEA of susy gauge theories in four dimensions from matrices
    \cite{DV0909,IMO0911}
   (\cite{IYosh1507, Itoy2016, LeFl2006}, for review).
The critical behavior \cite{AD9505,APSW9511,KY9712} of these LEEA translates into that of matrix models. 
In these notes, we shed light upon a few statistical field theoretic aspects, 
which are phase structure and nonvanishing $\beta$ functions.
   
    We follow the notation of section three in \cite{IYosh2411}: 
   The partition function is defined by 
   \begin{equation}
   	Z_n = \int \prod_{i=1}^N \de \alpha_i 
   	\prod_{i\neq j}   |\ex^{\im \alpha_i} - \ex^{\im \alpha_j}|
   	\exp \left\{-\frac{2N}{\lambda} \sum_{i} U_n(\alpha_i) \right\}, 
   \end{equation}
   \begin{equation}
   	(U_n(\alpha)) = - \cos \alpha- \sum_{k=2}^{n} \tau_k \cos k \alpha. 
   \end{equation}
     In \cite{IYosh2411}, the complete determination of phase has been given for $n=2$.
  Let us observe that the presence of  transition lines and hence the qualitative structure of the phase diagrams in this case can be reproduced 
  by knowing how the three different patters of shape for the classical potential evolve with the decrease of the parameter $\lambda$ that measures the strength of  
   repulsion among eigenvalues and at the same time couples to the operator carrying  lowest dimension. We begin with discussing this in the next section
    and show how the reasoning  can be generalized and analyzed when the potential contains higher orders. This will be  illustrated for  $\cos n\alpha$, $n \leq 3$.
In section 3, we derive a set of  $\beta$ functions for $n=1, 2$. 
We find that nowhere is there a vanishing vector, which tells us  the nature of 3rd order phase transition
 rather than the 2nd order, in conformity with the values of susceptibility exponent.

% % % % % % % % % % % % % % % % % % % % % % % % % % % % % % % % % % % % % % % % % %
% % % % % % % % % % % % % % % % % % % % % % % % % % % % % % % % % % % % % % % % % %
\section{phase structure from classical potential}
% % % % % % % % % % % % % % % % % % % % % % % % % % % % % % % % % % % % % % % % % %
% % % % % % % % % % % % % % % % % % % % % % % % % % % % % % % % % % % % % % % % % %
The potential is given by 
\begin{align}
W(\alpha) = \frac{1}{\lambda} U_n(\alpha) ,~~~~~
U_n(\alpha) = - \cos \alpha- \sum_{k=2}^{n} \tau_k \cos k \alpha , 
\end{align}
As the potential $U_n(\alpha)$ is even, we pay our attention to  $0 \leq \alpha \leq \pi$. 

The derivative of $U_n(\alpha)$ with respect to $\alpha$ is
\begin{equation}
 U'_n(\alpha) %=  \sum_{k=1}^n k \tau_i \sin (k \alpha) 
 = \sin \alpha  \times (\text{$(n-1)$-th degree polynomial in $\cos \alpha$}). 
\end{equation} 
We see that $\alpha = 0, \pi$ are always local extrema. 
In addition, it can be seen that there are at most $(n-1)$ local extrema within the interval $0 < \alpha < \pi$. 

The parameter $\lambda$  measures a repulsion between eigenvalues 
through the Vandermonde determintant.
If $\lambda$ is sufficiently large, the eigenvalues should spread 
 through the entire circle independently of the details of the potential.
As $\lambda$ decreases, the eigenvalues accumulate around smaller local minima of the potential.
In this way, the change in the eigenvalue distributions with $\lambda$
  is qualitatively determined by the shape of the classical potential.

% % % % % % % % % % % % % % % % % % % % % % % % % % % % % % % % % % % % % % % % % %
\subsection{ $\cos \alpha, \cos 2\alpha$}
% % % % % % % % % % % % % % % % % % % % % % % % % % % % % % % % % % % % % % % % % %

The $n=1$ is the case considered in \cite{GW1980,Wadi1980}. 
The potential is 
\begin{equation}\label{U:n=1}
 U_1(\alpha) = - \cos \alpha, 
\end{equation}
which has a local minimum at $\alpha = 0$. 
As $\lambda$ becomes smaller, the eigenvalues accumulate around the minimum $\alpha = 0$. 
Therefore, transition from the 0-gap to 1-gap is expected. 
In fact, the  $0 \to 1$ transition takes place at $\lambda = 2$. 

\begin{figure}[h]
 \centering
     \includegraphics[width=0.7\columnwidth]{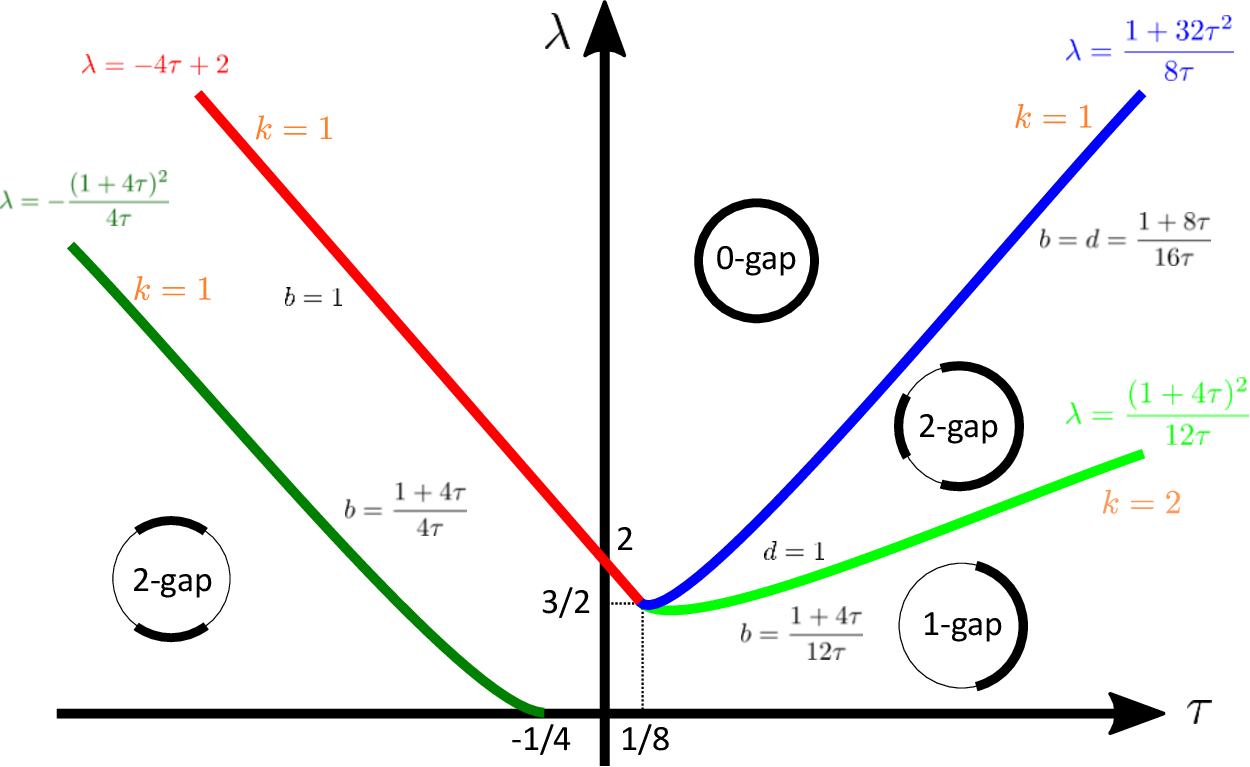}
     \caption{}
     \label{Fig:PD(n=2)}
 \end{figure}

The $n=2$ case was studied in \cite{Mand1990, IYosh2411} and the phase diagram 
 (seen in Fig.\ref{Fig:PD(n=2)}) was determined \cite{IYosh2411} 
 from the planar solution for the eigenvalue distribution.  
The classical potential has one parameter $\tau \equiv \tau_2$: 
\begin{equation}
 U_2(\alpha) = -\cos \alpha - \tau \cos 2 \alpha. 
\end{equation}

 %%%%%%%%%%%%%%%%%%%%%%%%%%%%%%%%%%
 %%%%%%%%%%%%%%%%%%%%%%%%%%%%%%%%%%　

\begin{figure}[h]
\centering
\begin{minipage}[b]{0.3\columnwidth}
    \centering
    \includegraphics[width=0.9\columnwidth]{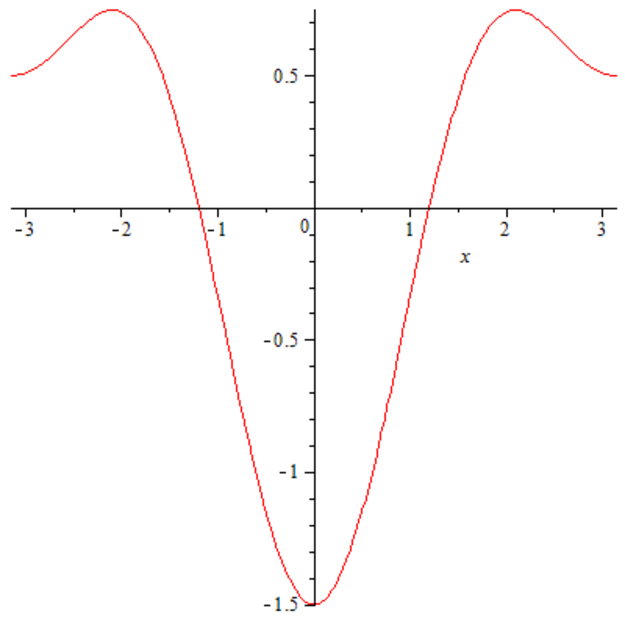}
    \caption{$\tau$= $\frac{1}{2}$,  
    $\alpha_0   = \frac{2\pi}{3}$.}
    \label{fig:U(1/2)}
\end{minipage}
\begin{minipage}[b]{0.3\columnwidth}
    \centering
    \includegraphics[width=0.9\columnwidth]{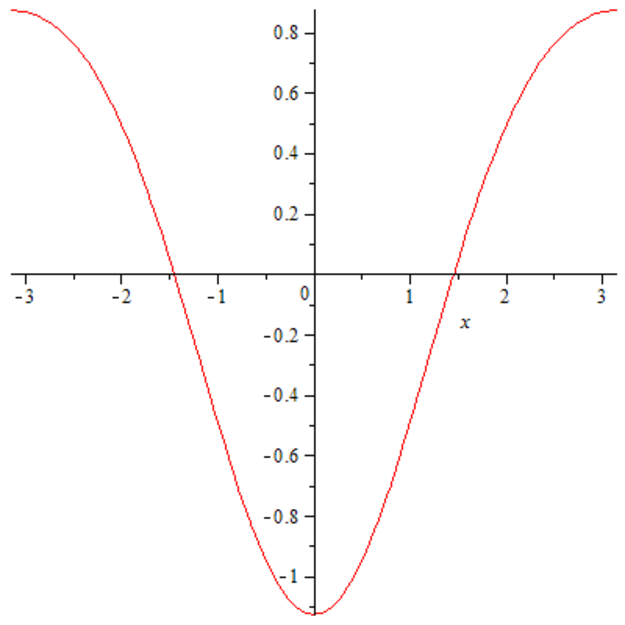}
    \caption{$\tau = \frac{1}{8}$.}
    \label{fig:U(1/8)}
\end{minipage}
\begin{minipage}[b]{0.3\columnwidth}
    \centering
    \includegraphics[width=0.9\columnwidth]{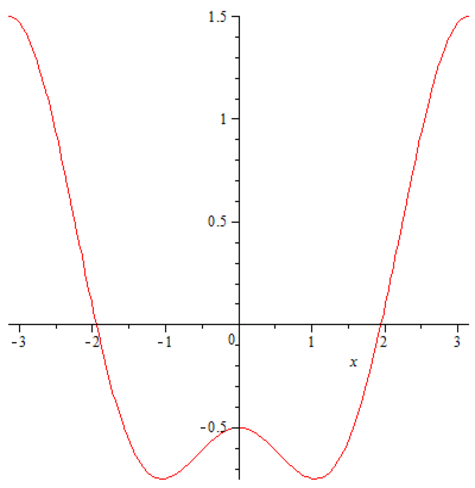}
    \caption{$\tau = -\frac{1}{2}$,  $\alpha_0= \frac{\pi}{3}$.}
    \label{fig:U(-1/2)}
\end{minipage}
\end{figure}

Depending upon the value of $\tau$, the shape of this potential changes 
(See, Fig.\ref{fig:U(1/2)}, Fig.\ref{fig:U(1/8)}, Fig.\ref{fig:U(-1/2)}.).
There are three different cases to consider. 

For large $\lambda$, the distribution ought to be zero gap for any case. 
As $\lambda$ decreases, the way the number of gaps changes differs for each case.
This and the statement in what follows is  seen clearly from Fig. \ref{Fig:PD(n=2)}.
 
 \noindent
 case (i) $\tau > \frac{1}{4}$:  $0 \Rightarrow 2 \Rightarrow 1$\\
There exist two local minima at $\alpha = 0$ and $\pi$ and two local maxima at $\alpha = \pm \alpha_0$. 
If $\lambda$ decreases, the eigenvalues accumulate around the local minima $\alpha = 0$ and $\pi$.  This is two-gap phase.   
At sufficiently small $\lambda$, all eigenvalues gather around the minimum $\alpha = 0$. Hence, one-gap phase for small $\lambda$ and we obtain $0 \to 2 \to 1$ transition. 
 
 \noindent
case (ii) $|\tau| \leq  \frac{1}{4}$ :    $0 \Rightarrow 1$ \\
There exist a local minimum at $\alpha=0$ and a local maximum at $\alpha = \pi$. 
In this case, the minimum is only at $\alpha=0$ and we obtain $0 \to 1$ transition.
 
\noindent
 case (iii) $\tau < -\frac{1}{4}$: $0 \Rightarrow 1 \Rightarrow 2$ \\
There exist  two local maxima at $\alpha = 0$, $\pi$ and two local minima at $\alpha = \pm \alpha_0$.      If $\lambda$ is large enough, eigenvalues go over the barrier. This is one-gap phase.  
As $\lambda$ gets smaller, the eigenvalues are distributed around each of the two minima.  
This is a two-gap phase and $0 \to 1 \to 2$ transition is expected. 

While it is not possible to determine the curves of the transition lines and the triple point $(1/8,3/2)$,
  the qualitative structure of the phase diagram is in fact obtained from
  the above classical reasoning alone.
 It is this simple point that we want to illustrate further.
 
%%%%%%%%%%%%%%%%%%%%%%%%%%%%%%%%%%
%%%%%%%%%%%%%%%%%%%%%%%%%%%%%%%%%%

% % % % % % % % % % % % % % % % % % % % % % % % % % % % % % % % % % % % % % % % % %
\subsection{ $\cos 3 \alpha$}
% % % % % % % % % % % % % % % % % % % % % % % % % % % % % % % % % % % % % % % % % %

It should be possible to determine the qualitative structure of the phase diagram from the potential.

The potential is given by 
\begin{equation}
 U_3(\alpha) = -\cos \alpha - \tau \cos 2 \alpha - \sigma \cos 3 \alpha, 
\end{equation}
where we set $\tau_1 = \tau$ and $\tau_2 = \sigma$. 
Let us see how the shape of the potential changes with the addition of 
 $\cos(3\alpha)$ term to $U_2(\alpha)$.  
For each of the five cases (I), (II), (III), (IV), (V), in what follows, 
 we first derive the behavior of $U_3(\alpha)$ with varying $\sigma$ 
 without relying on detailed calculations 
 and then we analyze the potential in detail to verify the results.
 For simplicity, we limit ourselves to the cases in which $\tau \geq 0$ and $\sigma \geq 0$. 

Note that  
at $\alpha = \frac{\pi}{6}, \frac{\pi}{2}, \frac{5\pi}{6}$, we have $\cos(3\alpha) = 0$ and, therefore,   
 the potential does not depend on $\sigma$ at these points, 
\begin{equation}
 U_3\left(\frac{\pi}{6}\right) = -\frac{\sqrt{3} + \tau}{2},~~~~~
 U_3\left(\frac{\pi}{2}\right) = \tau,~~~~~
 U_3\left(\frac{5\pi}{6}\right) = \frac{\sqrt{3} - \tau}{2},
\end{equation}
while 
at $\alpha = 0, \frac{\pi}{3}, \frac{2\pi}{3}, \pi$, the derivative of $\cos(3\alpha)$ vanishes
 and, therefore, the gradient of $U_3(\alpha)$ does not depend on $\sigma$ at these points. 

The extrema of $U_3(\alpha)$ are determined by 
\begin{equation}\label{U3'}
 U'_3(\alpha) = \sin\alpha ( 1-3\sigma + 4\tau \cos \alpha + 12 \sigma \cos^2 \alpha) = 0. 
\end{equation}
We see that $\alpha =0, \pi$ are always extrema which do not depend on $\tau$ and $\sigma$.  
The other two solutions to eq. \eqref{U3'} are  
\begin{equation}
 \alpha_{\pm} = \cos^{-1} \left(-\frac{\tau}{6\sigma} \left(
 1 \pm \sqrt{1 + \frac{3\sigma(3\sigma-1)}{\tau^2}}
 \right)\right), 
\end{equation} 
provided that  the following conditions have to be fulfilled: 
\begin{equation}\label{condition1}
 1 + \frac{3\sigma(3\sigma-1)}{\tau^2} \geq 0,
\end{equation}
and
\begin{equation}\label{condition2}
 -1 \leq \cos \alpha_{\pm} \leq 1.  
\end{equation}
Note that, besides $\alpha=0$ and $\pi$, there can be at most two other extrema.

We consider the following five cases with different $\tau$.

\noindent
(I) Case 1: $0 \leq \tau \leq \frac{1}{4}$

\begin{figure}[h]
\centering
    \includegraphics[width=1\columnwidth]{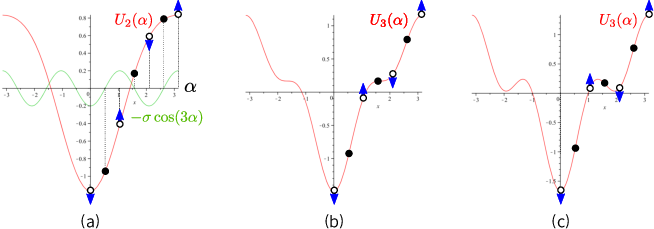}
    \caption{$\tau = \frac{1}{6}$. 
    The dots represent the points where $\cos(3\alpha)=0 $ and
    the circles represent the points where $(\cos(3\alpha))' = 3\sin(3\alpha) = 0$, 
    $(0 \leq \alpha \leq \pi)$.  
    These points move in the direction indicated by the arrow as $\sigma$ increases. 
    (a) $\sigma = 0$, 
    (b) $\sigma = \frac{3+2\sqrt{2}}{18}$, (c) $\sigma = \frac{1}{2}$.}
    \label{pot1}
\end{figure}

As shown in Fig.\ref{pot1}(a),  $U_3(\alpha ; \sigma = 0) = U_2(\alpha)$ has 
 no extrama except for those at $\alpha =0$ and $\pi$. 
%As can be seen from the figure of $-\cos(3\alpha)$, 
As $\sigma$ increases, the gradient at  $\alpha = \frac{\pi}{2}$ 
 decreases and eventually becomes negative
 while those at $\alpha = \frac{\pi}{6}$, $\frac{5\pi}{6}$ become steeper.
It is easy to observe that 
 the potential changes from (a) to (b) and then to (c) in Fig \ref{pot1}.
 
Let us examine the potential $U_3(\alpha)$ analytically. 
At $\sigma = \frac{1}{6}(1 + \sqrt{1-4\tau^2})$, 
 we have $\alpha_+ = \alpha_- = \cos^{-1} (-\frac{\tau}{6\sigma})$ 
 (See Fig.\ref{pot1}(b)).
If $\sigma$ is further increased, 
there will be a local maximum at  $\alpha_-$ and a  local minimum at $\alpha_+$ (See Fig.\ref{pot1}(c)).
 
%%%%%%%%%%%%%%%%%%%%%%%%%%%%%%%%%%
%%%%%%%%%%%%%%%%%%%%%%%%%%%%%%%%%%

\noindent
(II) Case 2: $\frac{1}{4} < \tau < \frac{1}{2}$

\begin{figure}[h]
\centering
    \includegraphics[width=1\columnwidth]{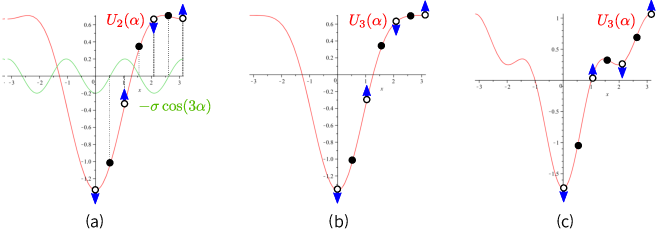}
    \caption{$\tau = \frac{1}{3}$. 
    (a) $\sigma = 0$, 
    (b) $\sigma = \frac{1}{27}$, (c) $\sigma = \frac{2}{5} > \frac{3 + \sqrt{5}}{18}$.}
    \label{pot2}
\end{figure}

In the case of $\tau > \frac{1}{4}$, $U_3(\alpha;\sigma=0) = U_2(\alpha)$ has a local maximum 
at $\alpha_0(\sigma = 0) = \cos^{-1} (-\frac{1}{4\tau})$ as shown in Fig.\ref{pot2}(a). 
As long as $\tau < \frac{1}{2}$, we have $\alpha_0 (\sigma = 0) > \frac{2\pi}{3}$. 
This implies that,  as $\sigma$ increases,  this local  maximum point $\alpha_0(\sigma)$ approaches $\pi$ because the gradient of $-\cos(3\alpha)$ is positive 
 in $\frac{2\pi}{3} < \alpha < \pi$. 
When $\alpha_0(\sigma) = \pi$, there are no extrema other than $\alpha=0$ and $\pi$. 
Simultaneously, the point $\alpha = \pi$ changes from a local minimum to maximum.
As $\sigma$ increases further and surpasses a certain value,  
 the gradient around $\alpha = \frac{\pi}{2}$ will change to negative, 
 yielding additional local maximum and minimum. 
Hence, it can be seen that the potential deforms as Fig \ref{pot2} (a)-(c).

The local maximum $\alpha_0(\sigma)$ is given by  
\begin{equation}
 \alpha_0(\sigma) = \begin{cases}
  \cos^{-1} (-\frac{1}{4\tau}) ~~~~~ \sigma = 0 \\
  \alpha_- ~~~~~ \sigma > 0
 \end{cases}. 
\end{equation} 
At $\sigma = \frac{1}{9}(4\tau - 1)$ for $\tau \leq \frac{2}{5}$, 
at $\sigma = \frac{\tau}{6}$ for $\tau > \frac{2}{5}$,
this local maximum point becomes $\alpha_-  = \pi$ (See Fig.\ref{pot2}(b)). 
At $\sigma = \frac{1}{6} (1 + \sqrt{1-4\tau^2})$, we have $\alpha_+ = \alpha_- = \cos^{-1} (-\frac{\tau}{6\sigma})$. 
When $\sigma$ becomes larger than that, 
 there will be a local maximum at  $\alpha_-$ and a  local minimum at $\alpha_+$
 (See Fig.\ref{pot2}(c)).

%%%%%%%%%%%%%%%%%%%%%%%%%%%%%%%%%%
%%%%%%%%%%%%%%%%%%%%%%%%%%%%%%%%%%

\noindent
(III) Case 3: $\tau = \frac{1}{2}$

\begin{figure}[h]
\centering
    \includegraphics[width=1\columnwidth]{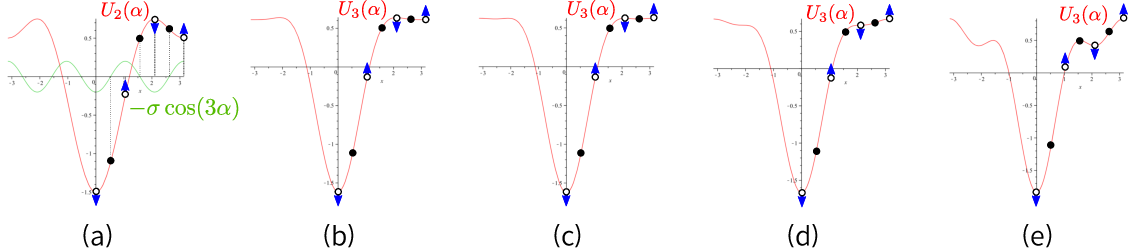}
    \caption{$\tau = \frac{1}{2}$. 
    (a) $\sigma = 0$, 
    (b) $\sigma = \frac{1}{9}$, (c) $\sigma = \frac{1}{8}$, (d) $\sigma = \frac{1}{6}$, (e) $\sigma = \frac{1}{3}$.}
    \label{pot3}
\end{figure}

For any $\sigma$, $\alpha=\frac{2\pi}{3}$ is the extremum of $U_3(\alpha)$. 
When $\sigma=0$,  we have $U_3(\frac{2\pi}{3}) > U_3(\pi)$. 
As $\sigma$ increases, $U_3(\frac{2\pi}{3})$ decreases and $U_3(\pi)$ increases.
At a certain $\sigma$, %$\neq 0$,
the local minimum at $\alpha=\pi$ should turn to  local maximum, and  
another local  minimum will appear in $\frac{2\pi}{3} < \alpha < \pi$. 
There have to be a $\sigma$  where $U_3(\frac{2\pi}{3}) = U_3(\pi)$. 
$U_3(\frac{2\pi}{3})$ keeps decreasing, and 
the local maximum at $\alpha=\frac{2\pi}{3}$ will turn to  the local minimum. 
Since $\alpha=0$ is always the minimum, 
there should be  a local maximum between them. 
Hence, it can be seen that the potential deforms as Fig. \ref{pot3} (a)-(e).

If $0 < \sigma < \frac{1}{6}$, there exists only one local maximum at $\alpha = \frac{2\pi}{3}$
 (See Fig.\ref{pot3}(a)-(c)). 
Another local minimum appears at $\alpha_+$ when $\sigma>\frac{1}{9}$.   
In particular, at $\sigma = \frac{1}{8}$, the values of two local maximum become equal,  
 $U_3(\frac{2\pi}{3}) = U_ 3(\pi)$  (See Fig. \ref{pot3}(c)). 
At $\sigma= \frac{1}{6}$, we  have $\alpha_- = \alpha_+$ %which is the inflection point 
  (See Fig.\ref{pot3}(d)). 
Finally,  in the case of $\sigma > \frac{1}{6}$,  
 there will be a local maximum at  $\alpha_-$ and a  local minimum at $\alpha_+$
  (See Fig.\ref{pot3}(e)). 

%%%%%%%%%%%%%%%%%%%%%%%%%%%%%%%%%%
%%%%%%%%%%%%%%%%%%%%%%%%%%%%%%%%%% 

\noindent
(IV) Case 4: $\frac{1}{2} < \tau \leq 1$

\begin{figure}[h]
\centering
    \includegraphics[width=1\columnwidth]{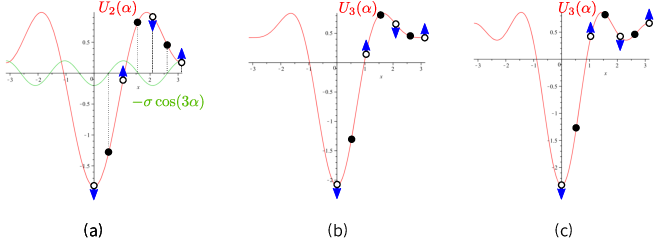}
    \caption{$\tau = \frac{5}{6}$. 
    (a) $\sigma = 0$, 
    (b) $\sigma = \frac{7}{27}$, (c) $\sigma = \frac{1}{2}$.}
    \label{pot4}
\end{figure}

At $\sigma = 0$, the local maximum point $\alpha_0(\sigma = 0)$ 
 is located within $\frac{\pi}{3} < \alpha < \frac{2\pi}{3}$.
As $\sigma$ increases,  $\alpha_0(\sigma)$ approaches $\frac{\pi}{3}$
 because the gradient of $-\cos(3\alpha)$ is negative in $ \frac{\pi}{3}<\alpha<\frac{2\pi}{3}$. 
The local minimum at $\alpha = \pi$ changes to  local maximum at a certain $\sigma$, 
 and another local minimum appears in  $\frac{2\pi}{3} < \alpha < \pi$. 
Although $U_3(\pi) < U_3(\alpha_0)$ at $\sigma=0$,  
 there should exist a $\sigma$ at which $U_3(\pi) = U_3(\alpha_0)$. 
This is  because  we have $\alpha_0 \to \frac{\pi}{3}$ in the limit $\sigma \to \infty$ and 
$U_3(\pi) = 1 - \tau + \sigma  > U_3(\frac{\pi}{3}) = \frac{-1+\tau}{2} + \sigma$ always follows. 
Hence, it can be seen that the potential deforms as Fig. \ref{pot4} (a)-(c).

In fact, at $\sigma \leq \frac{1}{9}(4\tau - 1)$, there is a local maximum at $\alpha_-$. 
At $\sigma > \frac{1}{9} (4\tau -1)$, a local minimum additionally appears at $\alpha_+$.

%%%%%%%%%%%%%%%%%%%%%%%%%%%%%%%%%%
%%%%%%%%%%%%%%%%%%%%%%%%%%%%%%%%%% 

\noindent
(V) Case 5: $\tau > 1$

\begin{figure}[h]
\centering
    \includegraphics[width=1\columnwidth]{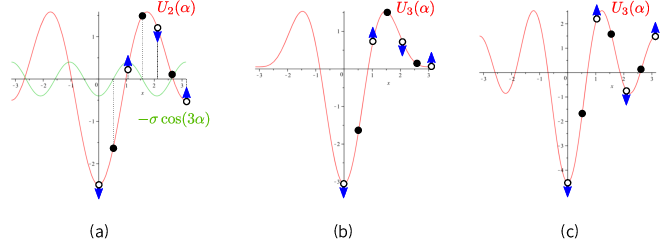}
    \caption{$\tau = \frac{3}{2}$. 
    (a) $\sigma = 0$, 
    (b) $\sigma = \frac{5}{9}$, (c) $\sigma = 2$.}
    \label{pot5}
\end{figure}

As can be seen in Fig. \ref{pot5},
the behavior of the local extremum $\alpha_0(\sigma)$ and 
 around $\alpha = \pi$ is similar to Case 4 . 
The difference is that $U_3(\pi) < U_3(\frac{\pi}{3})$ and $U_3(\alpha_-)$  is always the maximum value
 for any $\sigma \geq 0$. 
 
The extrema of $U_3(\alpha)$ is the same as Case 4.
We can confirm that $U_3(\alpha_-)$ takes always the maximum value in $0 \leq \alpha \leq \pi$. 

%%%%%%%%%%%%%%%%%%%%%%%%%%%%%%%%%%
%%%%%%%%%%%%%%%%%%%%%%%%%%%%%%%%%% 

\subsection{phase diagram}

\begin{figure}[h]
\centering
    \includegraphics[width=0.9\columnwidth]{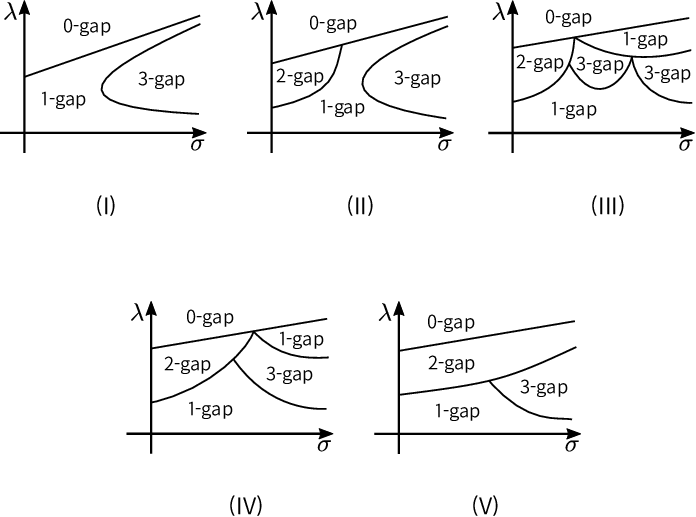}
    \caption{}
    \label{PD}
\end{figure}

Let us estimate the phase diagram expected from the shape of $U_3(\alpha)$. 
With $\tau$ fixed, we consider the phase diagram with respect to $\sigma$ and $\lambda$. 
For sufficiently large $\lambda$, the repulsion also becomes large and 
the eigenvalues spread throughout the entire space 
 and the model is in a 0-gap phase regardless of the potential.
 As $\lambda$ decreases, the shape of the potential begins to affect the eigenvalue distribution
and the eigenvalues should gradually accumulate around the local minima. 
The resulting phase diagrams are shown in Fig. \ref{PD} for each case.

%%%%%%%%%%%%%%%%%%%%%%%%%%%%%%%%%%
%%%%%%%%%%%%%%%%%%%%%%%%%%%%%%%%%%

\section{beta-function}

In this section, we investigate the beta-functions. 
Let
\begin{equation}
 w_k \equiv \lim_{N \to \infty} 
 \frac{1}{N} \left\langle \sum_{i=1}^N \cos (k\alpha_i) \right\rangle, 
 ~~~~~ k = 1, 2,\cdots, n. 
\end{equation}
It follows that $0 \leq w_k \leq 1$.
By definition, it is obvious that $w_k \leq 1$. 
The point  $\alpha=0$ is the minimum of the potential ($\tau_k > 0$) and
it implies that the eigenvalue density is maximized around this point.  
In its vicinity,  $\cos (k \alpha) > 0$ and, therefore, we see that $w_k\geq 0$. 

Let us introduce the scale parameter $a$ by 
\begin{equation}
w_k =  \ex^{-a^2 \sigma_k}, 
\end{equation}
where $\sigma_k$ correspond to $n$ kinds of string tensions. 
The beta function for $a$ is given by
\begin{equation}
 \beta_{k} \equiv a \frac{\de \tau_k}{\de a}= \sum W^{-1}_{k\ell}  \de_a w_{\ell},
 ~~~~~ \tau_1 \equiv \lambda, 
\end{equation} 
where $W^{-1}$ is the inverse matrix of
\begin{equation}
 W_{k \ell} =  \frac{\partial  w_k}{\partial \tau_{\ell}}, 
\end{equation}
and
\begin{equation}
 \de_a w_k %\equiv a \frac{\de w_k}{\de a} = .
 = 2 w_k \ln w_k. 
\end{equation}

Let us examine the GWW model ($n=1$ case). 
There is only one coupling constant $\lambda$. 
The free energy in the large $N$ limit is given by 
\begin{equation}
 F_{n=1} = \lim_{N \to \infty} \frac{\ln Z_{n=1}}{N^2} = 
 \begin{cases}
 \frac{1}{\lambda^2}  ~~~~~ \lambda \geq 2,\\
 \frac{2}{\lambda} + \frac{1}{2}\ln \frac{\lambda}{2} - \frac{3}{4} 
 ,~~~~~ 0 < \lambda \leq 2. 
 \end{cases}
\end{equation}
We can obtain 
\begin{equation}
w_1 =  -\frac{\lambda^2}{2} \frac{\partial F_{n=1} }{\partial \lambda} = 
\begin{cases}
 \frac{1}{\lambda}, \\
 1 -\frac{\lambda}{4}. 
\end{cases}
\end{equation}
The beta function is 
\begin{equation}
 \beta_1 =  2 w_1 \left(\frac{\partial  w_1}{\partial \lambda}\right)^{-1} \ln w_1 = 
\begin{cases}
 2 \lambda \ln \lambda \\
 2(4 - \lambda) \ln \frac{4}{4-\lambda}. 
\end{cases} .
\end{equation}
At the transition point $\lambda = 2$, $\beta_1 = 4 \ln 2$. 

In the $n=2$ model,  
\begin{align}
 &w_1 \equiv \lim_{N \to \infty} \frac{1}{N} \left\langle \sum_i^N \cos \alpha_i \right\rangle
  = w_{\lambda} - \tau w_{\tau}, \\
 &w_2 \equiv \lim_{N \to \infty} \frac{1}{N} \left\langle \sum_i^N \cos 2\alpha_i \right\rangle
 = w_{\tau}. 
\end{align}
where 
\begin{align}
&w_{\lambda} %= -\frac{\lambda^2}{2N^2} \frac{\partial \ln Z }{\partial \lambda}
 = -\frac{\lambda^2}{2} \frac{\partial F_{n=2} }{\partial \lambda}, \\
&w_{\tau} %= \frac{\lambda}{2N^2} \frac{\partial \ln Z}{\partial \tau}
 = \frac{\lambda}{2} \frac{\partial F_{n=2}}{\partial \tau}, 
\end{align}
and $F_{n=2}$ is the free energy of $n=2$ model. 

In the 0-gap phase, the free energy is given by 
\begin{equation}
 F_{n=2}^{(0)} = \frac{1}{\lambda^2}(1 + 2 \tau^2), 
\end{equation}
from which, we obtain
\begin{equation}
 w_1 = \frac{1}{\lambda}, ~~~~~
 w_2 = \frac{2\tau}{\lambda} .  
\end{equation}
and
\begin{equation}
 W ^{-1} = -\mat{\lambda^2 & 0 \\ \tau \lambda & - \lambda/2}
\end{equation}
Hence, 
\begin{equation}
 \beta_1 = 2 \lambda \ln \lambda, ~~~~~
 \beta_2 = 2 \tau \ln 2 \tau. 
\end{equation}

\begin{figure}[h]
\centering
    \includegraphics[width=0.9\columnwidth]{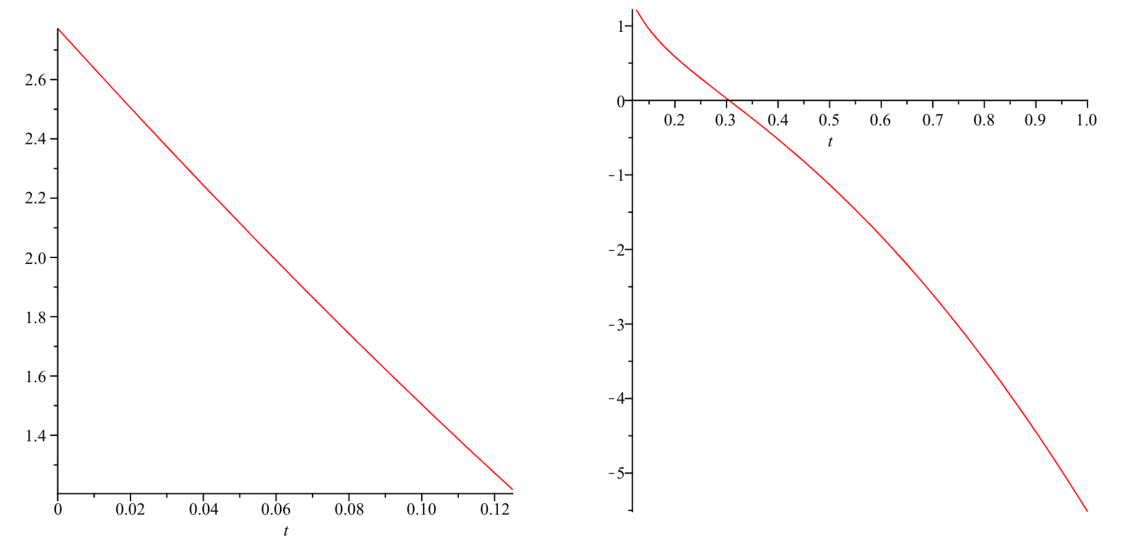}
    \caption{The beta function for $\lambda$ on the transition line.}
    \label{b1}
\end{figure}

\begin{figure}[h]
\centering
    \includegraphics[width=0.9\columnwidth]{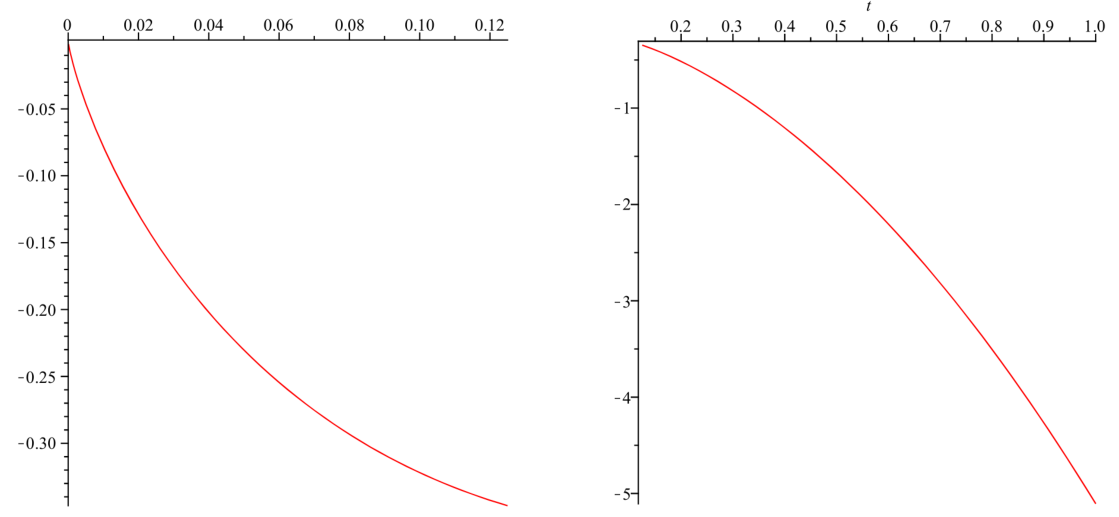}
    \caption{The beta function for $\tau$ on the transition line.}
    \label{b2}
\end{figure}

In the 1-gap phase, the free energy is given by 
\begin{align}
 F_{n=2}^{(1)} = &-\frac{1}{\lambda^2} \biggl\{
   18 \tau^2 b^4 -b^3(40 \tau^2 +10 \tau) + b^2(40 \tau^2 + 32\tau + 1) -b(16\tau^2 + 20 \tau + 4)
   \biggr\} \nonumber \\
   & -\frac{1}{\lambda} \biggl\{
   (6\tau b^2 -b(1+4\tau))\ln b  +(\ln 2)(-12\tau b^2 +b(2+ 8\tau)) - 3 \tau b^2 + b(1+4\tau)
   \biggr\} \nonumber \\
   &+ \ln 2.
\end{align}
From this free energy, we obtain 
\begin{equation}\label{w_1}
w_1 = \frac{1}{216} 
 \frac{(1+4\tau)^3 -18\tau \lambda (1-8\tau) - ((1+4\tau)^2 - 12 \tau \lambda)^{3/2})}{\tau^2\lambda}, 
\end{equation}
and 
\begin{equation}\label{w_2}
w_2 = \frac{1}{432} 
 \frac{-(1-4\tau)(1+4\tau)^3 +((1+4\tau)^2 - 12\tau\lambda)^{3/2}(1-4\tau) + 18\tau\lambda (1-3\tau\lambda +8\tau^2)}{\tau^3 \lambda}.
\end{equation}
The beta function for this model can be determined, and 
its behavior on the phase transition line is shown in Fig. \ref{b1} and Fig. \ref{b2}. 
It can be seen that the two beta functions never both be vanishing at the same time.

%%%%%%%%%%%%%%%%%%%%%%%%%%%%%%%%%%
%%%%%%%%%%%%%%%%%%%%%%%%%%%%%%%%%%
%%%%%%%%%%%%%%%%%%%%%%%%%%%%%%%%%%
%%%%%%%%%%%%%%%%%%%%%%%%%%%%%%%%%%

%%%%%%%%%%%%%%%%%%%%%%%%%%%%%%%%%%%%%%%%%%%%%%%%%%%%%
%%%%%%%%%%%%%%%%%%%%%%%%%%%%%%%%%%%%%%%%%%%%%%%%%%%%%
\section*{Acknowledgments}
We thank Takahiro Nishinaka for continuing discussion on this subject. 
The work of H.I. and R.Y. is supported in part by JSPS KAKENHI (23K03393, 23K03394).
%%%%%%%%%%%%%%%%%%%%%%%%%%%%%%%%%%%%%%%%%%%%%%%%%%%%%
%%%%%%%%%%%%%%%%%%%%%%%%%%%%%%%%%%%%%%%%%%%%%%%%%%%%%

%%%%%%%%%%%%%%%%%%%%%%%%%%%%%%%%%%%%%%%%%
%\bibliographystyle{arxiv}
%\bibliography{multicritical}
%%%%%%%%%%%%%%%%%%%%%%%%%%%%%%%%%%%%%%%%%

%%%%%%%%%%%%%%%%%%%%%%%%%%%%%%%%%%%%%%%%%%%%%%%%%%%%%

%%%%%%%%%%%%%%%%%%%%%%%%%%%%%%%%%%%%%%%%%%%%%%%%%%%%%

%%%%%%%%%%%%%%%%%%%%%%%%%%%%%%%%%%%%%%%%%%%%%%%%%%%%
%%%%%%%%%%%%%%%%%%%%%%%%%%%%%%%%%%%%%%%%%%%%%%%%%%%%%
\end{document}